\documentclass[aps,pre,
preprint,
12pt,
showpacs,
floatfix,
]{revtex4-1}
\usepackage{amsmath}    

\usepackage{graphicx}   
\usepackage{color}
\usepackage[hang]{subfigure}
\usepackage{hyperref}
\newcommand{\jun}{junction }
\newcommand{\juns}{junctions }
\newcommand{\Jos}{Josephson }
\newcommand{\elli}{elliptic }
\newcommand{\ann}{annular }

\begin{document}
\title[R.Monaco \textit{et al.}]{Confocal Annular Josephson Tunnel Junctions}
\author{Roberto Monaco}
\affiliation{CNR-ISASI, Institute of Applied Sciences and Intelligent Systems ''E. Caianello'', Comprensorio Olivetti, 80078 Pozzuoli, Italy}
\email[Corresponding author e-mail:]{r.monaco@isasi.cnr.it}
\date{\today}

\begin{abstract}
The physics of Josephson tunnel junctions drastically depends on their geometrical configurations and here we show that also tiny geometrical details play a determinant role. More specifically, we develop the theory of short and long confocal annular Josephson tunnel junctions in the presence of an in-plane magnetic field of arbitrary orientations. The behavior of a circular annular Josephson tunnel junction is then seen to be simply a special case of the above result. For junctions having a normalized perimeter less than one the threshold curves are derived and computed even in the case with trapped Josephson vortices. For longer junctions a numerical analysis is carried out after the derivation of the appropriate motion equation for the Josephson phase. We found that the system is modeled by a modified and perturbed sine-Gordon equation with a space dependent effective Josephson penetration length inversely proportional to the local junction width. Both the fluxon statics and dynamics are deeply affected by the non-uniform annulus width. Static zero-field multiple-fluxon solutions exist even in presence of a large bias current. The tangential velocity of a traveling fluxon is not determined by the balance between the driving and drag forces due to the dissipative losses. Furthermore, the fluxon motion is characterized by a strong radial inward acceleration which causes electromagnetic radiation concentrated at the ellipse equatorial points.
\end{abstract}
\maketitle

\section{Introduction}

The static \cite{SUST15} and dynamic \cite{JLTP16} properties were recently studied for Elliptic Annular \Jos Tunnel Junctions (EAJTJs) in the presence of a uniform in-plane magnetic field. An EAJTJ consists of two superconducting elliptic annuli coupled by a thin dielectric layer. An elliptic annulus, by definition, has a constant width and is implemented by drawing two closed curves {\it parallel} to a master ellipse, with constant but opposite offsets. The internal and external boundaries of such an annulus are not ellipses, but more complex curves \cite{http} (that will be given later). When the ellipse eccentricity vanishes, then the EAJTJ reduces to the well-known circular annular \Jos tunnel junction ideal for experimental tests of the perturbation models developed to take into account the dissipative effects in the propagation with no collisions of sine-Gordon kinks \cite{davidson, dueholm,hue}. In presence of an in-plane magnetic field, circular AJTJs were also recognized to be the ideal device to investigate both the statics and the dynamics of sine-Gordon solitons in a spatially periodic potential  \cite{gronbech, ustinov,PRB98,wallraf3}. There is, however, another configuration that generalizes the circular AJTJ: it is given by the Confocal Annular \Jos Tunnel Junction (CAJTJ) which is delimited by two ellipses having the same foci; for such geometry the annulus width is not constant. Therefore, as computer numerical control programmers know, elliptic and confocal annuli, although apparently similar, are quite different objects. In this work we develop the theory for both \textit{short} and \textit{long} CAJTJs in presence of an arbitrary in-plane magnetic field and will show that, despite the minor geometrical differences, their properties are markedly different from those of EAJTJs. More specifically, it will turn out that the phenomenology of CAJTJs, due to their non-uniform width, is much richer than that of EAJTJs, in both absence and presence of an external magnetic field.

\noindent The paper is organized as follows. In the rest of this Section we state the problem by stressing the difference between elliptic and confocal AJTJs and introduce the mathematical notations and identities used in the paper. In next Section we consider small JTJs immersed in a uniform in-plane magnetic field and compute first the threshold curves for elliptic \juns having different ellipticity; later we extend the analysis to CAJTJs with possible \Jos vortices trapped in the \ann barrier. In Section III we derive the appropriate partial differential equation for an electrically long CAJTJ; later, in Section IV, we present numerical simulations concerning the fluxon(s) static and dynamic properties. The conclusions are drawn in Section V.

\subsection{Elliptic versus confocal annuli}

To clarify the difference between elliptic and confocal annuli, let us consider the master ellipse $x^2/a^2+y^2/b^2=1$ centered in the origin of a Cartesian coordinate system whose $X$ and $Y$ axes are directed, respectively, along the principal ellipse diameters $2a$ and $2b$. We define the axes ratio $\rho\equiv b/a$ and the eccentricity $e^2 \equiv 1-\rho^2$. If $a>b$ then the ellipse foci lie on the $X$-axis and it is possible to find two positive numbers, $c$ and $\bar{\nu}$, such that $a=c \cosh \bar{\nu}$ and $b=c \sinh \bar{\nu}$; then $\rho=\tanh\bar{\nu}$ and $c= \pm\sqrt{a^2-b^2}$ are the abscissae of the ellipse's foci. The master ellipse is described by the the parametric equations:
\vskip -10pt
\begin{equation}
\label{master}
\begin{cases}
x(\tau)=c\cosh\bar{\nu}\sin\tau \\
y(\tau)=c\sinh\bar{\nu}\cos\tau, \end{cases} 
\end{equation}

\noindent where $\tau$ is a parameter measured clockwise from the positive $Y$-axis such that $\tan\tau=\tan\theta\tanh\bar{\nu}$, where the polar angle $\theta$ is defined as $\theta \equiv \text{ArcTan}\, x/y$. When the ellipse tends to a circle of radius $r=a=b$, $\rho=\tanh\bar{\nu}=1$, then $\tau\to\theta$, $c\to0$ and $\bar{\nu}\to \infty$, while $c\cosh\bar{\nu}\approx c\sinh\bar{\nu}\to r$. For a segment of length $2a$, $b\to 0$, then $c\to a$ and $\bar{\nu}\to 0$. If $b>a$, then the foci lie on the $Y$-axis and $c$ is an imaginary number, while, with $\rho>1$, $\nu=\arctan\!\textrm{h}\,\rho= \arctan\!\textrm{h}\rho^{-1} -\hat{\imath} \pi/2$ is a complex number.

\noindent The parametric equation of the inner and outer boundaries of the  elliptic annulus with (constant) width $\Delta w$ are given, respectively, by \cite{http}: 
\vskip -10pt

\begin{equation}
\label{inner}
\begin{cases}
x_i(\tau)=\left[c\cosh\bar{\nu}-\Delta w \sinh\bar{\nu}/2\mathcal{Q}(\tau)\right]\sin\tau; \\
y_i(\tau)=\left[c\sinh\bar{\nu}-\Delta w \cosh\bar{\nu}/2\mathcal{Q}(\tau)\right]\cos\tau; \end{cases} 
\end{equation}
\vskip -10pt
\noindent and
\vskip -10pt
\begin{equation}
\label{outer}
\begin{cases}
x_o(\tau)=\left[c\cosh\bar{\nu}+\Delta w \sinh\bar{\nu}/2\mathcal{Q}(\tau)\right]\sin\tau; \\
y_o(\tau)=\left[c\sinh\bar{\nu}+\Delta w \cosh\bar{\nu}/2\mathcal{Q}(\tau)\right]\cos\tau; \end{cases} 
\end{equation}
\vskip 2pt

\begin{figure}[tb]
\centering
\includegraphics[width=10cm]{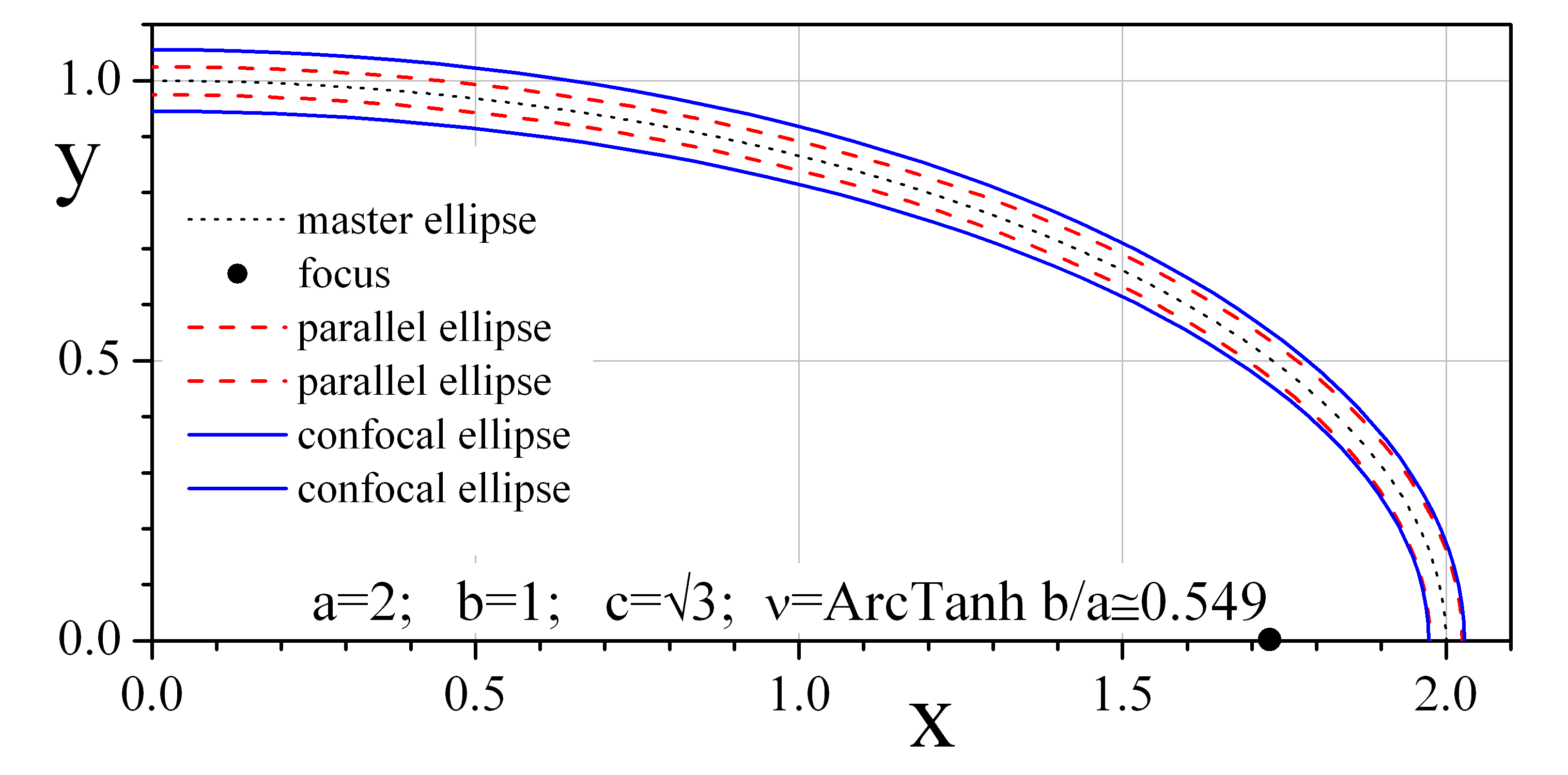}
\caption{(Color online) Illustration of the difference between a confocal annulus delimited by the two confocal ellipses (blue solid lines) and an elliptic annulus bounded by the two {\it parallel} ellipses (red dashed lines). The {\it parallel} ellipses are not ellipses - see Eqs.(\ref{inner}) and (\ref{outer}).}
\label{superpos}
\end{figure}

\noindent where $\mathcal{Q}^2(\tau) \equiv q^2(\bar{\nu},\tau)=\sinh^2\!\bar{\nu} \sin^2\tau +\cosh^2\!\bar{\nu} \cos^2\tau = \sinh^2\!\bar{\nu}+\cos^2\!\tau=\cosh^2\!\bar{\nu} - \sin^2\!\tau=(\cosh\!2\bar{\nu} + \cos\!2\tau)/2>1$. Very simply, the parametric equation of the inner and outer boundaries of the confocal annulus are, respectively,: 

\begin{equation}
\label{inner2}
\begin{cases}
x_i(\tau)=c\cosh\nu_i\sin\tau; \\
y_i(\tau)=c\sinh\nu_i\cos\tau; \end{cases} 
\end{equation}
\noindent and
\begin{equation}
\label{outer2}
\begin{cases}
x_o(\tau)=c\cosh\nu_o\sin\tau; \\
y_o(\tau)=c\sinh\nu_o\cos\tau. \end{cases} 
\end{equation}

\noindent where $(\nu_o+\nu_i)/2=\bar{\nu}$. The width of such annulus is a $\pi$-periodic function of $\tau$; in fact,:
$$\Delta w(\tau)\!=\!c\sqrt{(x_o-x_i)^2+(y_o-y_i)^2}\!=\! c\sqrt{(\cosh\!\nu_o-\cosh\!\nu_i)^2\sin^2\!\tau +(\sinh\!\nu_o-\sinh\!\nu_i)^2\cos^2\!\tau}.$$

\noindent If $\Delta\nu\equiv \nu_o-\nu_i<<1$, the expression of the width reduces to:

\vskip -10pt
\begin{equation}
\nonumber
\Delta w(\tau)=c\mathcal{Q}(\tau)\,\Delta\nu.
\end{equation}
\vskip -5pt

\noindent Its maximum value is $\Delta w_{max}=c\cosh\bar{\nu}\Delta\nu=a\Delta\nu$ at the ellipse poles, $\tau=m\pi$ ($m$ integer), while $\Delta w_{min}=c\sinh\bar{\nu} \Delta\nu=b\Delta\nu$ is the minimum value achieved at the equatorial points, $\tau=m\pi\pm \pi/2$. The width relative variation $(\Delta w_{max}-\Delta w_{min})/\Delta w_{min} =\coth \bar{\nu}-1$ diverges as $\bar{\nu} \to 0 $. Therefore, the discrepancy (or disparity) between confocal and elliptic AJTJs are more evident for eccentric geometries.  

\noindent In Figure~\ref{superpos} we plot the parametric equations in (\ref{master}), (\ref{inner}), (\ref{outer}), (\ref{inner2}) and (\ref{outer2}); to emphasize the subtle distinction between confocal and elliptic annuli we restricted the curves to the first quadrant, i.e., for $0\leq\tau\leq\pi/2$. For the master ellipse (dotted curve) we set $a=2$ and $b=1$ (resulting in $\rho=0.5$, $\bar{\nu}=\arctan\!\textrm{h} \rho\simeq0.549$ and $c=\sqrt{3}$); for the two confocal ellipses (blue solid lines) we choose $\Delta\nu=0.1\bar{\nu}$, while for the two parallel ellipses (red dashed lines) it was $\Delta w=b \Delta\nu$. With such choices the two annuli have the same width at the equatorial points. To keep going with the parallelism we can state that, as an elliptic annuli has a constant $\Delta w$, a confocal one has a constant $\Delta\nu$. Nevertheless, for $a=b$, both geometries reduce to the circular annulus.

\subsection{The planar elliptic coordinates}

\noindent We now introduce the (planar) confocal elliptic coordinates $(\nu,\tau)$, such that, for a positive $c$ value, any point $(x,y)$ in the $X$-$Y$ plane is given by $(c\cosh\nu\sin\tau, c\sinh\nu\cos\tau)$ with $\nu\geq0$ and $\tau\in[-\pi,\pi]$. The elliptic variable
$\tau$ has a domain $[-\pi,\pi]$ and plays a role similar to that of the polar angle in polar coordinates; The $\nu$ coordinates behaves as a radial variable and identifies confocal ellipses centered on the origin, that is, the ellipse in Eq.(\ref{master}) has the equation $\nu(\tau)= \bar{\nu}$. The line joining the foci $(\pm c ,0)$ corresponds to $\nu=0$. Notice that the polar coordinates could be considered to be a special case of the elliptic coordinates in the limit $c\to 0$ when the foci of the elliptic coordinates collapse to a point at the origin. In elliptic coordinates the elementary distance is $ds= \sqrt{dx^2+dy^2} =f(\nu,\tau)\sqrt{d\nu^2+ d\tau^2}$, where $f(\nu,\tau)=c\,q(\nu,\tau)$ is the so-called scale factor with $q^2(\nu,\tau)=\sinh^2\nu \sin^2\tau+\cosh^2 \nu \cos^2 \tau= \sinh^2\nu+\cos^2\tau=\cosh^2\nu - \sin^2\tau=(\cosh2\nu + \cos2\tau)/2$. Furthermore, the elementary surface element is $dS=dxdy=f^2d\nu d\tau$. A vector ${\bf H}$ applied at a point $(\nu,\tau)$ can be decomposed in its normal and tangential components, respectively, $H_\nu={\bf \hat{N}} \cdot {\bf {H}}$ and $H_\tau={\bf \hat{T}} \cdot {\bf {H}}$, were:

\begin{subequations}
\begin{eqnarray}
{\bf{\hat{N}}}\equiv&[\sinh\nu\sin\tau/q(\nu,\tau),\cosh\nu \cos\tau/q(\nu,\tau)], \label{nu} \\
{\bf \hat{T}}\equiv&\,[\cosh\nu \cos\tau/q(\nu,\tau), -\sinh\nu\sin\tau/q(\nu,\tau)], 
\label{tau}
\end{eqnarray}
\end{subequations}
 
\noindent are, respectively, the (outward) normal and (clockwise) tangent unit vectors to the ellipse passing at the point $(\nu,\tau)$; in different words, ${\bf\hat{N}}$ and ${\bf\hat{T}}$ form an orthonormal basis on two-dimensional vectorial space. Throughout the paper we will carry out the analysis assuming $a>b$; however, all the derived expressions will still be real when $a<b$, provided that $c$ is replaced by its imaginary counterpart, $\hat{\imath}\,c$ \cite{note1}. 

\section{Small junctions}

In Josephson's original description the quantum mechanical phase difference, $\phi$, across the barrier of a generic two-dimensional planar \Jos tunnel \jun is related to the magnetic field, ${\bf H}$, inside the barrier \cite{brian} through:
\vskip -10pt
\begin{equation}
\label{gra}{\bf \nabla} \phi =
\kappa{\bf H}\times {\bf u}_z ,
\end{equation}

\noindent in which ${\bf u}_z$ is a unit vector orthogonal to the \jun plane and $\kappa^{-1}\equiv \Phi_0/2\pi\mu_0 d_m$, where $\Phi_0$ is the magnetic flux quantum, $\mu_0$ the vacuum permeability, and $d_m$ the \jun \textit{magnetic} penetration depth \cite{wei,SUST13a}. 

\noindent With the in-plane magnetic field applied at an arbitrary angle ${\bar{\theta}}$ with the $Y$-axis, ${\bf {H}}=(H \sin\bar{\theta},H\cos\bar{\theta})$, in force of Eq.(\ref{gra}) the \Jos phase is $\phi(x,y)=\kappa H(y\sin\bar{\theta} -x\cos\bar{\theta})+\phi_0$, where $\phi_0$ is an integration constant. Passing to elliptic coordinates, it is:

$$\phi(\nu,\tau,\bar{\theta}) = \kappa H c(\sin\bar{\theta} \sinh\nu \cos\tau - \cos\bar{\theta} \cosh\nu \sin\tau) +\phi_0=$$

\begin{equation}
=\pi \frac{H}{\tilde{H}}\left[ \frac{\sin\bar{\theta} \sinh\nu}{q(\nu,\bar{\theta})} \cos\tau - \frac{\cos\bar{\theta} \cosh\nu}{q(\nu,\bar{\theta})} \sin\tau \right] + \phi_0=h \sin(\xi-\tau) + \phi_0,
\label{small}
\end{equation}

\noindent where $\tilde{H}(\nu,\bar{\theta})=\Phi_0/2\mu_0 d_m c\, q(\nu,\bar{\theta})$.  In Eq.(\ref{small}) we also introduced the adimensional magnetic field $h(\nu,\bar{\theta})=\pi H/\tilde{H}$ and the angle $\xi(\nu,\bar{\theta})$ such that $\sin\xi=\sin\bar{\theta} \sinh\nu/q(\nu,\bar{\theta})$ and $\cos\xi=\cos\bar{\theta} \cosh\nu/q(\nu,\bar{\theta})$. In passing, we observe that $\tan \xi=\tan\bar{\theta} \tanh\nu$; this implies that when $\theta(\tau)$ coincides with the field orientation $\bar{\theta}$, then $\xi=\tau$, i.e., for any $h$ and $\nu$ value, $\phi_0=\phi(\nu,\bar{\tau},\bar{\theta})$ where $\tan\bar{\tau}\equiv\tan\bar{\theta} \tanh \nu$.

\subsection{Small \elli junctions}

\noindent To begin with, we first consider a simply-connected planar \Jos tunnel \jun delimited by an ellipse of principal semi-axes $a=c\cosh\bar{\nu}<<\lambda_J$ and $b=2c\sinh\bar{\nu}$ in presence of a spatially homogeneous in-plane magnetic field. The tunnel currents flow in the $Z$-direction and the local density of the \Jos current in \elli coordinates can be expressed as \cite{brian}:
\vskip -15pt
\begin{equation}
\label{jj}J_J(\nu,\tau) =J_c(\nu,\tau) \sin \phi(\nu,\tau) ,
\end{equation}

\noindent where the maximum \Jos current density, $J_c$, generally speaking, depends on both $\nu$ and $\tau$ and is constant inside uniform barrier junctions. The \Jos current, $I_J$, through the barrier is obtained integrating Eq.(\ref{jj}) over the junction area, $A$; assuming that $J_c$ is constant over the junction area:

\begin{equation}
\label{IJ}
I_J=\int_A J_J dS=J_c \int_A \sin \phi\, dS.
\end{equation}

\noindent Inserting Eq.(\ref{small}) in Eq.(\ref{IJ}) and carrying out the calculations reported in the Appendix A, we get (see Eq.(\ref{AZ2})):
\vskip -6pt
\begin{equation}
\label{IJellipse}
I_J(h,\phi_0)= 2 J_c A \sin\phi_0  \frac{J_1(h)}{h}
\end{equation}

\noindent in which $A=\pi a b = \pi c^2 \sinh\bar{\nu}\cosh\bar{\nu}$ is the ellipse area and $J_n$ the $n$-th order Bessel function of the first kind. $I_J$ is largest when $\phi_0=\pm \pi/2$, so the magnetic diffraction pattern (MDP), $I_c(H,\bar{\theta})$, for a small \elli \jun is:
\vskip -6pt
\begin{equation}
\label{IcEllipse}
I_c(H,\bar{\theta})= \max_{\phi_0} I_J(H,\phi_0)=J_J A \left| \frac{J_{1}(\pi H/\bar{H})}{\pi H/2\bar{H}}\right|,
\end{equation}

\noindent where $\bar{H}(\bar{\theta})=\Phi_0/\mu_0 d_m L(\bar{\theta})$ with $L(\bar{\theta}) \equiv c\,q(\bar{\nu},\bar{\theta})$ is the junction characteristic field . It can be shown that $2L(\bar{\theta})$ is the length of the projection of the junction in the direction normal to the externally applied magnetic field; as expected, $2L(0)=2a$ and $2L(\pi)=2b$. Eq.(\ref{IcEllipse}), first reported by Peterson \textit{et al.} \cite{ekin} in 1990, generalizes the so called \textit{Airy pattern} of a circular junction \cite{barone} of radius $r=a=b$.

\subsection{Small confocal \ann junctions}

The MDP of a small CAJTJ can be readily computed from Eq.(\ref{AZ}) by setting the integration limits in $\nu'$ from $\nu_i$ to $\nu_o$, where $\nu_i$ to $\nu_o$ identify, respectively, the inner and outer ellipses delimiting the \jun area (see Eqs.(\ref{inner2}) and (\ref{outer2})); then, Eq.(\ref{IJ}) can be rewritten as:  
\vskip -15pt
\begin{equation}
\label{IJ2}
I_J(h,\phi_0)=J_c c^2\int_{\nu_i}^{\nu_o} \! \Im(\nu,\phi_0) d\nu,
\end{equation}

\noindent with $\Im(\nu,\phi_0)$ given by Eq.(\ref{Inu}). In the absence of trapped fluxons ($n=0$), inserting Eq.(\ref{prim}) in Eq.(\ref{IJ2}), it is:
\begin{equation}
\label{IcAnnElli0}
I_c(H,\bar{\theta}) = J_J \left|A_o \frac{J_{1}(\pi H/\bar{H_o})}{\pi H/2\bar{H_o}} - A_i \frac{J_{1}(\pi H/\bar{H_i})}{\pi H/2\bar{H_i}}\right|,
\end{equation}

\noindent where $A_{i,o}=\pi a_{i,o} b_{i,o}=\pi c^2 \sinh \nu_{i,o} \cosh \nu_{i,o}$, $\bar{H}_{i,o}(\bar{\theta})=\Phi_0/\mu_0 d_m L_{i,o}(\bar{\theta})$ and $L_{i,o}(\bar{\theta}) =(a_{i,o}^2\cos^2\bar{\theta} + b_{i,o}^2 \sin^2\bar{\theta})^{1/2}= c(\cosh^2\nu_{i,o} \cos^2\bar{\theta} +\sinh^2\nu_{i,o} \sin^2\bar{\theta})^{1/2}=c\,q(\nu_{i,o},\bar{\theta})$. For $n\neq0$ the numerical computation of the integral in Eq.(\ref{IJ2}) yields:
\begin{equation}
\label{IcAnnElli}
I_c(H,\bar{\theta}) \approx J_J \left|A_o \frac{J_{n+1}(\pi H/\bar{H_o})}{\pi H/2\bar{H_o}} - A_i \frac{J_{n+1}(\pi H/\bar{H_i})}{\pi H/2\bar{H_i}}\right|,
\end{equation}

\noindent The approximation gets better and better as either $\left|n\right|$ and/or $H$ increases or when both $\nu_{i}$ and $\nu_{o}$ get larger and larger, meaning that the confocal annulus tends to a ring. 
\vskip 15pt
%
%
%

\subsection{Narrow small confocal \ann junctions}

An exact expression for the MDP of a small CAJTJ can be obtained for arbitrary winding number when the annulus width is infinitesimal, i.e., when $\Delta \nu=\nu_{o} -\nu_{i} <<1$. In this case Eq.(\ref{IJ2}) reduces to :
\vskip -10pt
$$I_J(h,\phi_0)=J_c c^2 \Im(\bar{\nu}) \Delta\nu,$$

\noindent where $\bar{\nu}\equiv (\nu_o+\nu_i)/2$. Inserting Eq.(\ref{Inu}) and considering that the annulus area is $\Delta A=\pi c^2 \cosh2\bar{\nu}\Delta\nu$, we have:
\vskip -15pt
\begin{equation}
\label{IcNarrowConfAnn}
I_c(h,\bar{\theta}) = J_c \Delta A  \left| \frac{\sin2\bar{\xi}}{\cosh\!2\bar{\nu} } \frac{J_{n-2}(h) - J_{n+2}(h)}{2} + J_{n}(h) +  \frac{\cos2\bar{\xi}}{\cosh\!2\bar{\nu} } \frac{J_{n-2}(h) + J_{n+2}(h)}{2} \right|,
\end{equation}


\noindent where $\bar{\xi}(\bar{\theta})=\xi(\bar{\nu},\bar{\theta})$ and, as for \elli junctions, $h=\pi H/\bar{H}$ with $\bar{H}(\bar{\theta})=\Phi_0/\mu_0 d_m L(\bar{\theta})$. It is possible to demonstrate that, in limit $\nu_i \to \nu_o$, Eq.(\ref{IcAnnElli0}) reduces to Eq.(\ref{IcNarrowConfAnn}) with $n=0$. As soon as $\bar{\nu}$ exceeds the unity, then $\cosh2\bar{\nu}>>1$; therefore, for slightly eccentric CAJTJs, Eq.(\ref{IcNarrowConfAnn}) simplifies to: 
\vskip -10pt
$$I_c(H,\bar{\theta}) \approx J_c \Delta A  \left| J_n \left(\frac{\pi H}{\bar{H}}\right) \right|.$$

\begin{figure}[tb]
\centering
\subfigure[ ]{\includegraphics[width=7cm]{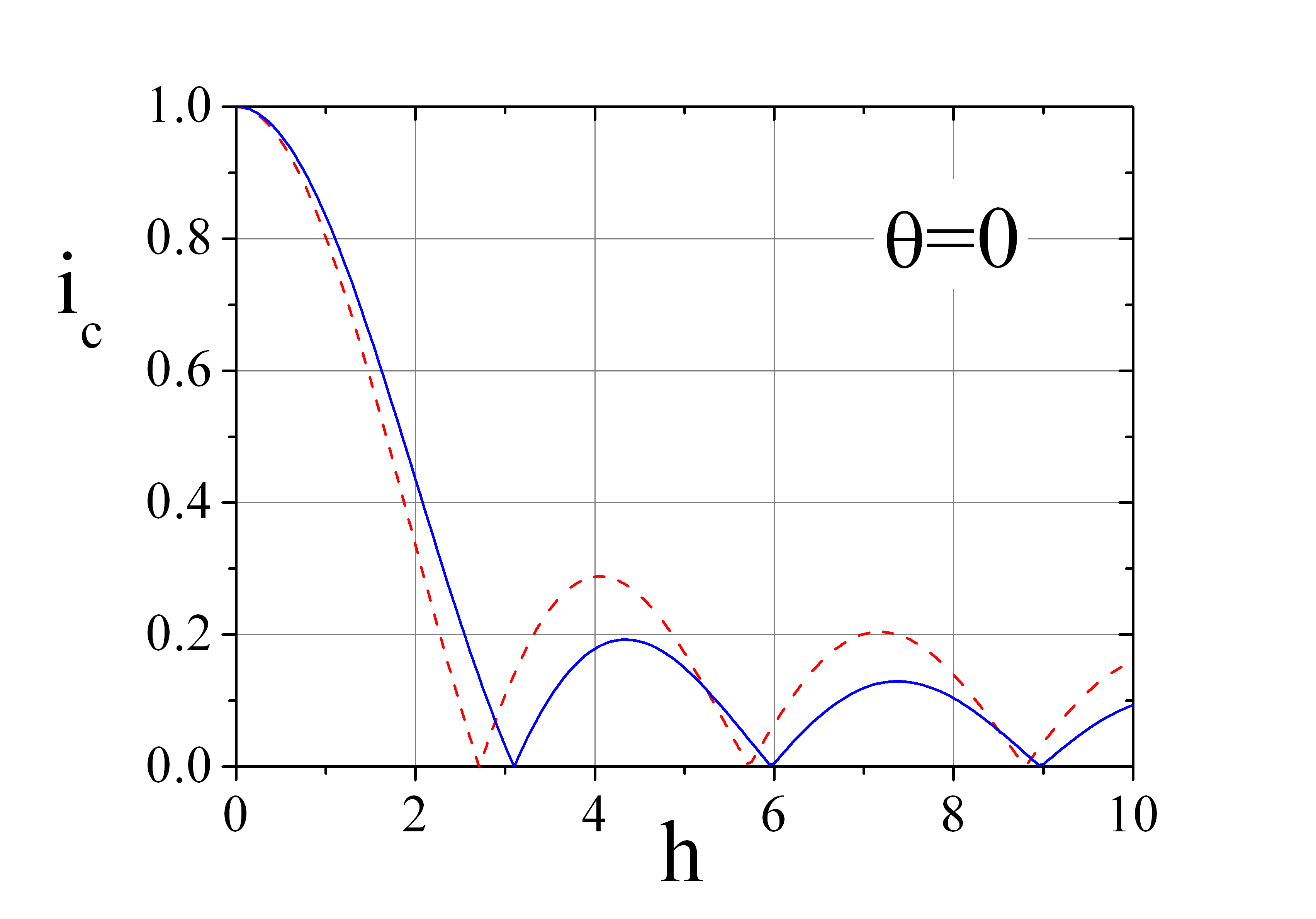}}
\subfigure[ ]{\includegraphics[width=7cm]{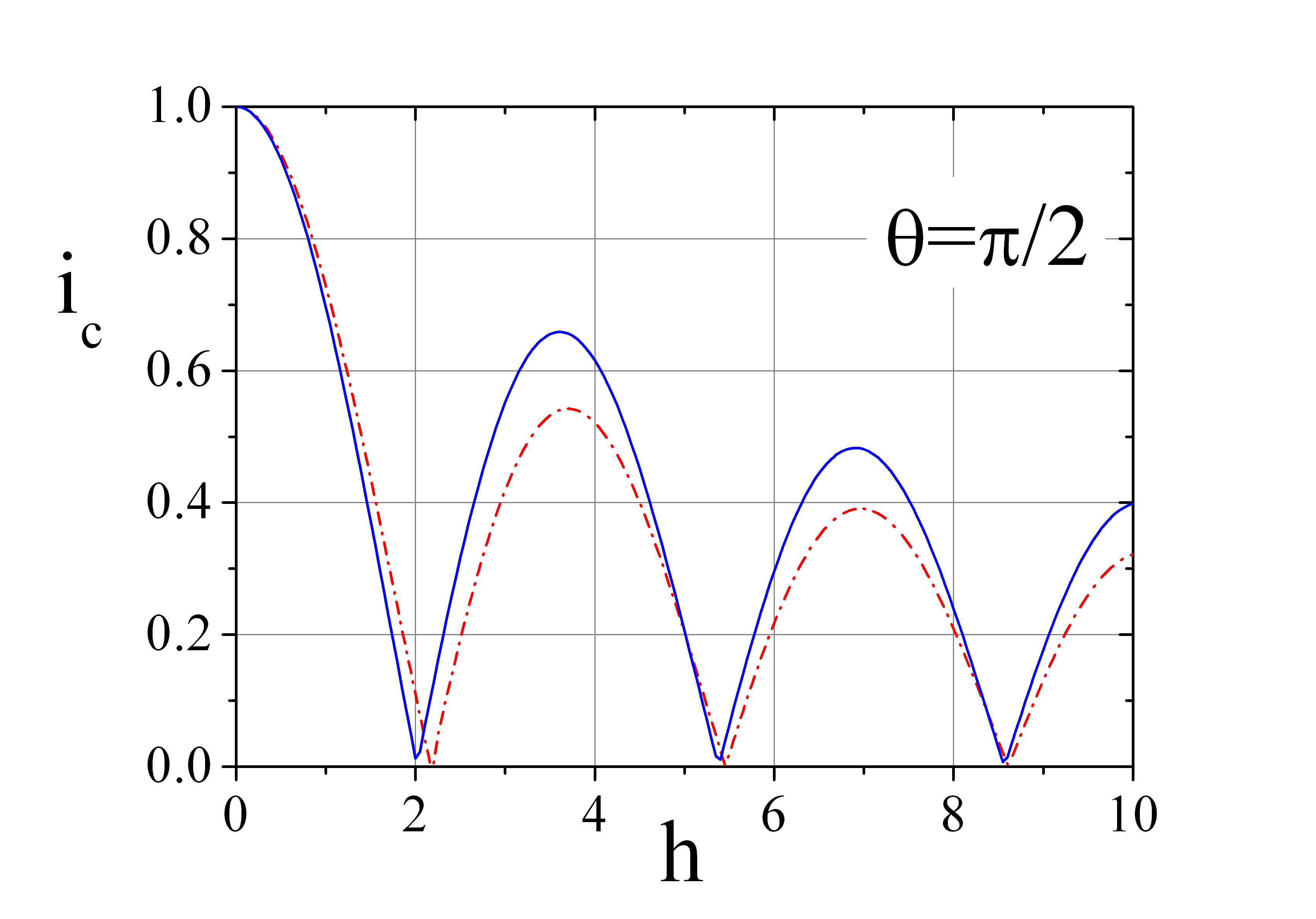}}
\caption{Comparison of the MDPs of a small and narrow confocal (blue solid line) and elliptic (red dashed line) annular junction having $\rho=0.5$ and $n=0$ for two orientations of the applied in-plane magnetic field: (a) $\bar{\theta}=0$ and (b) $\bar{\theta} =\pi/2$. Since $i_c(-h)=i_c(h)$, we only show the dependence for positive field values.}
\label{PatternComparisons}
\end{figure}

\noindent This equation has been already reported, but for the more restrictive case of narrow ring-shaped junctions \cite{br96}. It is worth to point out that in Eqs.(\ref{IcEllipse}), (\ref{IcAnnElli0}), (\ref{IcAnnElli}) and (\ref{IcNarrowConfAnn}) the $\bar{\theta}$ dependence is hidden in the characteristic field ${\bar{H}}$. In Figure~\ref{PatternComparisons}(a) we compare the MDPs of small and narrow confocal (blue solid line) and elliptic (red dashed line) annular junctions having $\rho=0.5$ (as those drawn in Figure~\ref{superpos}) in presence of a magnetic field parallel to the $b$-axes ($\theta=0$); in Figure~\ref{PatternComparisons}(b) the same comparison is carried out for $\theta=\pi/2$. Eq.(\ref{IcNarrowConfAnn}) was used for the CAJTJ, while the expression for the EAJTJ was taken from Ref.\cite{SUST15}. We observe that, despite the tiny difference in the geometrical configurations, there are significant quantitative discrepancies in the $i_c(h)$ dependence. The disparity increases with the system eccentricity. It is also evident that for a CAJTJ in a uniform field the minima in the magnetic pattern are not integer multiples of the first one, although they are (almost) equally spaced, the separation between two contiguous minima being about $\pi$.

\section{Long one-dimensional CAJTJs} 

In this section we derive the partial differential equation (PDE) for the \Jos phase of a confocal AJTJ having the foci in $(\pm c,0)$ [and $\bar{\nu}=(\nu_o+\nu_i)/2$] in presence of a spatially homogeneous in-plane magnetic field of arbitrary orientation, $\bar{\theta}$. The total tunnel current density is given by:

$$J_Z=J_c\sin \phi +\frac {\Phi_0}{2\pi R}\phi_t ,$$

\noindent where the second term in the right side takes into account the quasi-particle tunnel current assumed to be ohmic, i.e., $R$ is the voltage independent quasi-particle resistance per unit area. The subscripts on $\phi$ denote partial derivatives. By combining the previous equations with Maxwell's equations, one obtains a $2+1$ non-linear PDE for $\phi$ \cite{barone}:

\begin{equation}\label{psge}
\lambda_J^2 \left(1+\frac{\beta}{\omega_p} \frac{\partial}{\partial t}\right) \nabla^2 \phi - \frac 1{\omega_{p}^2}\phi_{tt}-\sin \phi =\frac{\alpha}{\omega_p} \phi_t,
\end{equation}
\noindent where $\lambda_J^2={\Phi_0}/{2\pi \mu_0J_cd_j}$ and $\omega _{p}^2={2\pi J_c}/{\Phi_0 c_s}$, $d_j$ being \jun \textit{current} penetration depth \cite{wei,SUST13a} and $c_s$ the specific junction capacitance. It is well known that the parameter $\lambda_J$, called the \Jos penetration length of the junction, gives a measure of the distance over which significant spatial variations of the phase occur, in the time independent configuration; the plasma frequency $\omega_p/2\pi$ represents the oscillation frequency of small amplitude waves. Further, we can introduce the parameter $\overline{c}\equiv \omega_p\lambda_J=1/\sqrt{\mu_o d_j c_s}$ which gives the velocity of light in the barrier and is called Swihart velocity \cite{Swihart}. In the last equation the $\alpha$ and $\beta$ terms take into account, respectively, the quasi-particle shunt loss and the surface losses in the superconducting electrodes. Eq.\ref{psge} is called Perturbed sine-Gordon Equation (PSGE). Because of its local form, it is quite general and holds for planar \juns of any geometrical configuration. On the \jun boundary the continuity of the induction field is provided by \cite{pagano}:

\begin{equation}
\label{gra2}
\left(1+\frac{\beta}{\omega_p} \frac{\partial}{\partial t}\right){\bf \nabla} \phi =\kappa{\bf H}^{ext}\times {\bf u}_z ,
\end{equation}

\noindent where ${\bf H}^{ext}$ is the external field that, in general, is given by the sum of an externally applied field, ${\bf H}$, and the self-field, ${\bf H}^{cur}$, generated by the current flowing in the junction. Using the elliptic coordinates, Eqs.(\ref{psge}) and (\ref{gra2}) become, respectively:

\begin{equation}\label{psgeElliptic}
\frac{\lambda_J^2 }{f^2} \left(1+\frac{\beta}{\omega_p}\frac{\partial}{\partial t}\right) \left(\frac{\partial^2 \phi}{\partial \tau^2} + \frac{\partial^2 \phi}{\partial \nu^2} \right)- \frac 1{\omega
_{p}^2}\phi_{tt}-\sin \phi =\frac{\alpha}{\omega_p} \phi_t
\end{equation}
\noindent and 
\begin{equation}
\label{gra3}
\frac{1}{f} \left(1+\frac{\beta}{\omega_p}\frac{\partial}{\partial t}\right) \left( \frac{\partial \phi}{\partial \nu}, \frac{\partial \phi}{\partial \tau}\right)=	\kappa(H_\tau^{ext},-H_\nu^{ext}),
\end{equation}

\noindent with $\phi=\phi(\nu,\tau,t)$. In the small width approximation, $w_{max}<< \lambda_J$, the \Jos phase does not depends on $\nu$ and the system becomes one-dimensional, $\phi= \phi(\tau,t)$. Furthermore, the scale factor becomes $f(\bar{\nu},\tau)=c\, \mathcal{Q}(\tau)$, where $\mathcal{Q}^2(\tau) \equiv q^2(\bar{\nu},\tau)=\sinh^2\!\bar{\nu} \sin^2\tau +\cosh^2\!\bar{\nu} \cos^2\tau = \sinh^2\!\bar{\nu}+\cos^2\!\tau=\cosh^2\!\bar{\nu} - \sin^2\!\tau=(\cosh\!2\bar{\nu} + \cos\!2\tau)/2>1$. At last the elementary arc is $ds=c 	\mathcal{Q}(\tau)d\tau$. Following Benabdallah \textit{et al.} \cite{Benabdallah}, we can apply the averaging operator $\frac{1}{\Delta \nu} \int_{\nu_i}^{\nu_o} d\nu$ on Eq.(\ref{psgeElliptic}) and obtain:	
\vskip -10pt
\begin{equation}
\left[\frac{\lambda_J}{c\,\mathcal{Q}(\tau)}\right]^2 \left(1+\frac{\beta}{\omega_p}\frac{\partial}{\partial t}\right) \left[\frac{\partial^2 \tilde{\phi}}{\partial \tau^2} + \frac{1}{\Delta \nu} \left( \left.\frac{\partial \phi}{\partial \nu}\right|_{\nu=\nu_o} -\left. \frac{\partial \phi}{\partial \nu}\right|_{\nu=\nu_i} \right) \right]- \frac 1{\omega_{p}^2}\tilde{\phi}_{tt}-\sin \tilde{\phi} = \frac{\alpha}{\omega_p} \tilde{\phi}_t,
\label{psge3}
\end{equation}

\noindent where:

$$\tilde{\phi}(\tau,t)=\frac{1}{\Delta \nu}\int_{\nu_i}^{\nu_o} \phi(\nu,\tau,t)d\nu$$

\noindent and we assumed $\sin\bar{\phi}\simeq\sin\phi$. According to Eqs.(\ref{gra3}) and (\ref{psge3}), the exact knowledge of the tangential components of the external field allows the determination of the proper boundary conditions. With the in-plane magnetic field applied at a generic angle $\bar{\theta}$ with the $Y$-axis, ${\bf {H}}=(H \sin\bar{\theta}, H\cos\bar{\theta})$, recalling Eqs.(\ref{nu}) and (\ref{tau}), we have:

\vskip -5pt
\begin{subequations}
\begin{eqnarray}
H_\nu(\nu,\tau)={\bf \hat{N}} \cdot {\bf {H}}= &  H \left(\sin\bar{\theta}\sinh\nu\sin\tau + \cos\bar{\theta}\cosh\nu\cos\tau \right)/{q(\nu,\tau)}, \label{NormField} \\
H_\tau(\nu,\tau)={\bf \hat{T}} \cdot {\bf {H}}= & H  \left(\sin\bar{\theta}\cosh\nu\cos\tau -\cos\bar{\theta}\sinh\nu\sin\tau\right)/ {q(\nu,\tau)}, \label{TangField}
\end{eqnarray}
\end{subequations}

\noindent It turns out that $\left({\bf {\nabla}}\! \times \! {\bf {H}} \right)_z = \frac{1}{f^2} \left[\frac{\partial (f H_\nu)}{\partial \tau} - \frac{\partial (f H_\tau)}{\partial \nu} \right]=0$ as expected for a spatially homogeneous magnetic field that is irrotational. 
\noindent The self-field induced on the annulus boundaries by a distributed bias current, $I$, can be computed by applying the Ampere's circuital law along the inner and outer \jun perimeters: in the former case, $H_\tau^{cur}(\nu_i,\tau)=0$, because no current can flow through the annulus hole; in the latter case, the tangential field equals in amplitude the sheet current $j_z(\tau)=J_Z(\tau) w(\tau)$, i.e., $H_\tau^{cur}(\nu_o,\tau)=j_z(\tau)= J_Z(\tau) w(\tau)=c J_Z(\tau) \mathcal{Q}(\tau) \Delta\nu$ (it can be easily checked that the field circuitation along the outer perimeter equals the bias current, $\int_{S}J_Z(s)dS$). Therefore, the phase normal derivative on the outer and inner annulus boundaries are:


$$\left.\frac{\partial \phi}{\partial \nu}\right|_{\nu=\nu_o}= \kappa cq(\nu_o,\tau) H_\tau^e(\nu_o,\tau)= \kappa cq(\nu_o,\tau) [H_\tau(\nu_o,\tau)+H_\tau^{cur}(\nu_o,\tau)]=$$
$$=\kappa cH(\sin\bar{\theta}\cosh\nu_o\cos\tau-\cos\bar{\theta}\sinh\nu_o\sin\tau)+\kappa c^2 J_Z(\tau) \mathcal{Q}^2(\tau) \Delta\nu$$

\noindent and

$$\left.\frac{\partial \phi}{\partial \nu}\right|_{\nu=\nu_i}=\kappa cq(\nu_o,\tau) H_\tau(\nu_o,\tau) = \kappa c H(\sin\bar{\theta}\cosh\nu_i\cos\tau-\cos\bar{\theta}\sinh\nu_i\sin\tau).$$

\noindent By subtracting the last two expressions, to the first order, we get:

$$\frac{1}{\Delta \nu} \left( \left. \frac{\partial \phi}{\partial \nu}\right|_{\nu=\nu_o} - \left. \frac{\partial \phi}{\partial \nu}\right|_{\nu=\nu_i} \right)=
\kappa cH(\sin\bar{\theta}\sinh\bar{\nu}\cos\tau-\cos\bar{\theta}\cosh\bar{\nu}\sin\tau)+ \kappa c^2 J_Z(\tau) \mathcal{Q}^2(\tau)=$$
$$=\frac{c^2}{\lambda_J^2} h'(\sin\bar{\theta}\sinh\bar{\nu}\cos\tau -\cos\bar{\theta}\cosh\bar{\nu} \sin\tau)+ \frac{J_Z(\tau)}{J_c} \left[\frac{c	\mathcal{Q}(\tau)}{\lambda_J}\right]^2$$

\noindent where $h'=H/J_c c$ is the $\bar{\theta}$-independent normalized field for treating long CAJTJs and we have made use of the approximation $\kappa \approx 1/J_c \lambda_J^2$ valid for thick electrode junctions \cite{SUST13a}. Inserting the last expression in Eq.(\ref{psge3}) and normalizing the time to $\omega_p^{-1}$, we end up with the PDE of a long CAJTJ:

\begin{equation}
 \left[\frac{\lambda_J}{c\,\mathcal{Q} (\tau)}\right]^2 \left(1+\frac{\beta\partial}{\partial \hat{t}}\right) \tilde{\phi}_{\tau\tau} - \tilde{\phi}_{\hat{t}\hat{t}}-\sin \tilde{\phi} =\alpha\tilde{\phi}_{\hat{t}} - \gamma(\tau) + F_h(\bar{\theta},\bar{\nu},\tau),
\label{psge4}
\end{equation}
\vskip -1pt
\noindent where $\gamma(\tau)\equiv J_Z(\tau)/J_c$; for a bias current, $I$, uniformly distributed over the \jun area $A$, it is $J_z(s)=I/A$ and $\gamma(\tau)= \gamma_0 \cosh2\bar{\nu}/ 2\mathcal{Q}^2(\tau)$, where $\gamma_0\equiv I/J_cA$. Furthermore, 
\vskip -2pt
\begin{equation}
F_h(\bar{\theta},\bar{\nu},\tau)\equiv h'\Delta \frac{\cos\bar{\theta}\cosh\bar{\nu} \sin\tau-\sin\bar{\theta}\sinh\bar{\nu}\cos\tau }{\mathcal{Q}^2(\tau)}
\label{Fh}
\end{equation}
\noindent is a forcing term proportional to the applied magnetic field. $\Delta$ is a geometrical factor which sometimes has been referred to as the coupling between the external field and the flux density of the junction \cite{gronbech}. The non-linear PDE in Eq.(\ref{psge4}) is supplemented by the periodic boundary conditions \cite{PRB96}:
\vskip -10pt
\begin{subequations}
\begin{eqnarray} \label{peri1}
\tilde{\phi}(\tau+2\pi,\hat{t})=\tilde{\phi}(\tau,\hat{t})+ 2\pi n,\\
\tilde{\phi}_\tau(\tau+2\pi,\hat{t})=\tilde{\phi}_\tau(\tau,\hat{t}),
\label{peri2}
\end{eqnarray}
\end{subequations}

\noindent where $n$ is an integer number, called the winding number, corresponding to the algebraic sum of \Jos vortices (or fluxons) trapped in the \jun due to flux quantization in one of the superconducting electrodes. Once trapped the fluxons can never disappear and only fluxon-antifluxon ($F\bar{F}$) pairs can be nucleated. Eqs.(\ref{psge4}) can be classified as a perturbed and modified sine-Gordon equation in which the perturbations, as usual, are given by the system dissipation and driving fields, while the modification is represented by an effective local $\pi$-periodic \Jos penetration length, $\Lambda_J(\tau)\equiv \lambda_J/Q(\tau)= c \lambda_J \Delta \nu /\Delta\!W(\tau)$, inversely proportional to the annulus width. It is worth noting that the Swihart velocity is constant around the annulus; in fact, describing the transmission lines in terms of lumped elements, the self-inductance ${\cal L}$ and the capacitance ${\cal C}$ per unit length of the \Jos transmission line are \cite{Swihart}, respectively, ${\cal L}={\mu_0 d_j / \Delta\!W}$ and ${\cal C}={c_s \Delta\!W }$, hence, the phase velocity, $v_{p}={1/\sqrt{{\cal L}{\cal C}}}=\overline{c}$, is independent of the local width.
\subsection{Some comments}

Notably, the PDE of CAJTJ does not differ by that of a circular one \cite{PRB97}:
\vskip -10pt

\begin{equation}
\left(\frac{\lambda_J}{\bar r}\right)^2 \left(1+\frac{\beta\partial}{\partial \hat{t}}\right) \phi_{\theta\theta} - \phi _{\hat{t}\hat{t}}-\sin \phi = \alpha \phi_{\hat{t}} -\gamma_0 + \frac{H\Delta}{J_c r}\sin(\bar{\theta}-\theta),
\label{ring}
\end{equation}

\noindent provided that the space dependent scaling factor $c\mathcal{Q}$ is replaced by the ring mean radius $r$ and the tangential elliptic coordinate $\tau$ is changed into the polar angle $\theta$. In the limit of a vanishing eccentricity ($e^2\simeq0$ and $\cosh2\bar{\nu}>>1$), it is $2\mathcal{Q}^2(\tau)= \cosh2\bar{\nu}+\cos2\tau \approx \cosh2\bar{\nu}$, i.e., $\gamma(\tau)= \gamma_0$. It is easy to demonstrate that, in the limits $c\to0$ and $\nu\to\infty$, the elliptic coordinates $(\nu,\tau)$ reduce to polar coordinates $(r,\theta)$ and Eq.(\ref{Fh}) results in a sinusoidal forcing term. However, the forcing term corresponding to a uniform in-plane magnetic field is more convoluted in a CAJTJ. 
\vskip 5pt
\noindent Following Ref.\cite{SUST15}, the PSGE for an EAJTJ in a uniform in-plane field applied at an angle $\bar{\theta}$ can be rearranged as:

\vskip -10pt
\begin{equation}
\left(\frac{\lambda_J}{c\mathcal{Q}}\right)^2 \left(1+\frac{\beta\partial}{\partial \hat{t}}\right)\left[ \phi_{\tau\tau} + \frac{\sin\!2\tau} {2\mathcal{Q}^2} \,\phi_\tau \right]-\phi_{\hat{t}\hat{t}}-\sin\phi= \alpha\phi_{\hat{t}}- \gamma(\tau)  + F_h(\bar{\theta},\bar{\nu},\tau),
\label{EAJTJ}
\end{equation}

\noindent where now:
\vskip -10pt
\begin{equation}
F_h(\bar{\theta},\bar{\nu},\tau)=h'\Delta\frac{\sinh2\bar{\nu}}{2} \,\frac{\sin\bar{\theta} \cosh\bar{\nu}\cos\tau +\cos\bar{\theta}\sinh\bar{\nu}\sin\tau}{\mathcal{Q}^4(\tau)}.
\label{FhE}
\end{equation}

\noindent We observe that in Eq.(\ref{psge4}) valid for a CAJTJ the term proportional to $\phi_\tau$ is absent because the inner and outer annulus boundaries are confocal ellipses. Furthermore, the forcing term in Eqs.(\ref{Fh}) and (\ref{FhE}) are markedly different, despite the fact that confocal and elliptic AJTJs have quite similar shapes.

\section{Numerical simulations}
\vskip -10pt
\begin{figure}[b]
\centering
\includegraphics[width=8cm]{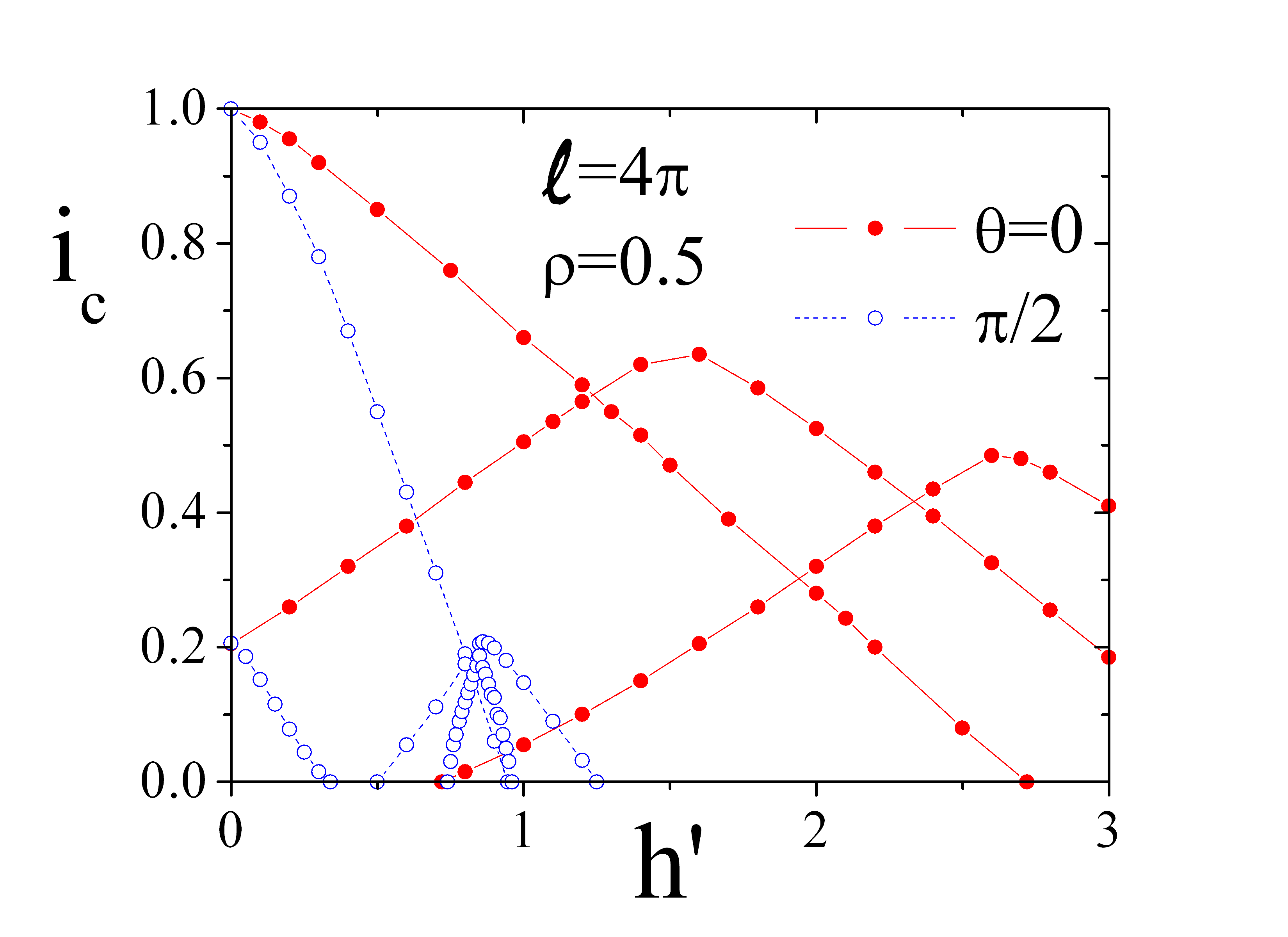}
\caption{(Color online) Numerical magnetic diffraction patterns, $i_c(h')$, of a long CAJTJ with $\ell=4\pi$, $\rho=0.5$ and $n=0$ for two values of the in-plane field orientation, $\theta=0$ and $\pi$. The magnetic field is normalized to $J_c c$.}
\label{4Pi}
\end{figure}

The commercial finite element simulation package COMSOL MULTIPHYSICS (\href{url}{www.comsol.com)} was used to numerically solve Eq.(\ref{psge4}) subjected to cyclic boundary conditions Eqs.(\ref{peri1}) and (\ref{peri2}). In all present calculations we set the damping coefficients $\alpha=0.1$ (weakly underdamped limit) and $\beta=0$, while keeping the current distribution uniform, i.e., $\gamma(\tau)= \gamma_0 \cosh2\bar{\nu}/ 2\mathcal{Q}^2(\tau)$. In addition, the coupling constant $\Delta$ was set to $1$. 

\subsection{The statics}
\vskip -10pt
\noindent Numerical solutions of Eq.(\ref{psge4}) have been carried out in the stationary, i.e., time-independent, state ($\phi_{\hat{t}}=0$) to compute the dependence on the magnetic field of the critical current of long CAJTJs.  Specifically, we have numerically computed the maximum value, $i_c$, of the zero-voltage current versus the normalized field amplitude, $h'$, setting $\ell=4\pi$, $\rho=0.5$ and $n=0$; our findings are shown in Figure~\ref{4Pi} for two different values of the in-plane field orientation, namely, $\theta=0$ (red closed dots and solid line) and $\pi$ (blue open dots and dashed line). Since $i_c(-h')=i_c(h')$, we only show the dependence for positive field values. As for small CAJTJs, the response to the external field is stronger when the field is perpendicular to the longer ellipse axes, although the modulation of the critical current is enhanced with the field parallel to the longer axes. The peculiarity of the long CAJTJs is the existence of static $F\bar{F}$ pairs for low field values which makes the critical current to be multiple-valued. In the figure we only plot the solutions corresponding to one pair. As the eccentricity is reduced the corresponding sub-lobe shrinks and eventually disappears; on contrary it gets larger as we increase the annulus normalized perimeter. The static $F\bar{F}$ solutions are the result of a potential barrier whose maxima are at $\tau=0$ and $\pi$ where the annulus is widest; this intrinsic barrier prevent the fluxon(s) to move until a threshold (depinning) value is reached by the bias current. The existence of a fluxon repelling (attracting) barrier induced by a widening (narrowing) \Jos transmission line was first found by Pagano \textit{et al.} \cite{pagano}; the barrier polarity is the same for fluxons and anti-fluxons. The main lobe of the magnetic diffraction pattern shows a linear dependence of the critical current on the external field; indeed, this feature is common to all long JTJs and can be erroneously interpreted as the signature of the full expulsion of the magnetic field from the junction interior (Meissner effect) that is not achievable in curved junctions \cite{SUST15}. Upon increasing the field amplitude, some ranges of magnetic field develop, in correspondence of the pattern minima, in which $i_c$ may assume two different values corresponding to different configurations of the \Jos phase inside the barrier \cite{owen}. In order to trace the different lobes of $i_c$ vs $h'$, it is crucial to start the numerical integration with a proper initial phase profile.

\subsection{Fluxon dynamics}
\vskip -10pt
In this subsection we analyze the dynamics of a magnetic flux quantum (current vortex) trapped in a current-biased long CAJTJ. We will limit to the case of no applied field; the consequences of an externally applied in-plane magnetic field will be considered in a future work. When fluxons are trapped in any annular \jun its zero-voltage critical current is considerably smaller; in addition, a stable finite-voltage current branch, called zero-field step (ZFS), appears in the junction current-voltage characteristic indicating that the bias current forces the fluxon(s) to travel along the annulus in the absence of collisions.
\begin{figure}[b]
\centering
\includegraphics[width=7cm]{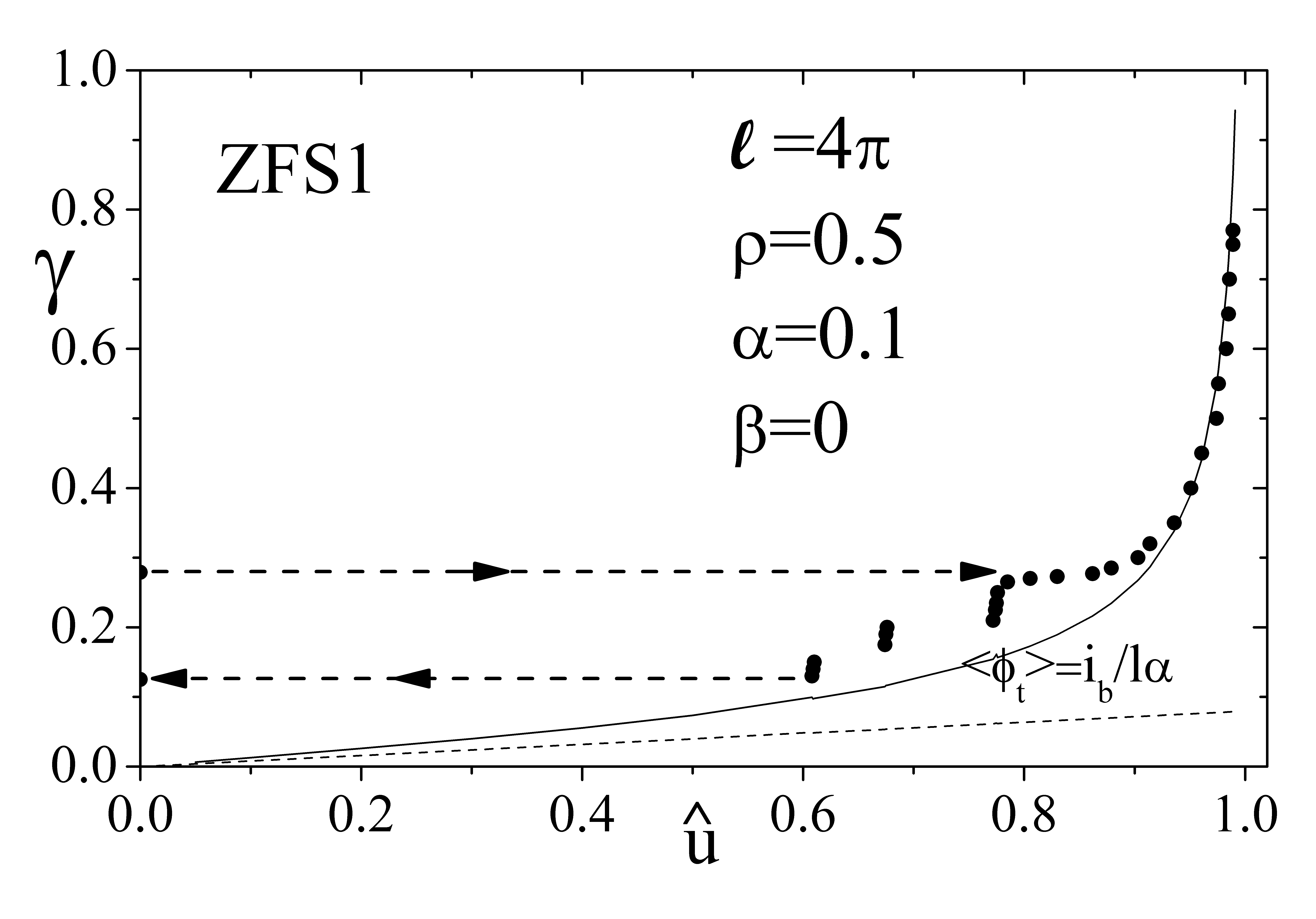}
\caption{The dots refer to the numerically computed profile of the first zero-field step for an CAJTJ.  Results are calculated integrating Eq.(\ref{psge4}) with $l=4\pi$, $\rho=1/2$, $\alpha=0.1$, $\beta=0$, $h=0$, and $n=1$. The solid line is the perturbative model expectation $\gamma(\hat{u})= 4\alpha/\pi \sqrt{\hat{u}^{-2}-1}$. The dashed line depicts the quasi-particle current $\gamma/\ell \alpha$.}
\label{ZFS1h0}
\end{figure}
\noindent The dots in Figure~\ref{ZFS1h0} show the numerically computed current-voltage (i.e., $\gamma$ versus $<\phi_{\hat{t}}>$) characteristic of a CAJTJ of normalized length $\ell=4\pi$ and axes-ratio $\rho=0.5$ with one trapped fluxon ($n=1$). Noticing that the mean voltage generated by a fluxon moving with velocity $\hat{u}$ is given by $V \propto <\phi_{\hat{t}}>=2\pi \hat{u}/\ell$, and since, from Eq.(\ref{psge}), $\gamma$ means a force, we can think of the plot in Figure~\ref{ZFS1h0} also as force-velocity characteristics. The normalized mean velocity, $\hat{u}$, was determined from the fluxon revolution period $\hat{T}$, as $\hat{u}=\ell/\hat{T}$ and was defined to be positive for fluxons rotating clockwise. The dashed line depicts the quasi-particle current $\gamma/\ell \alpha$. The ZFS profile results to be quite different from the perturbative model expectation $\gamma(\hat{u})= 4\alpha/\pi \sqrt{\hat{u}^{-2}-1}$ (solid line) appropriate to constant width long JTJs \cite{scott,JLTP16}. The main discrepancy is the appearance of a zero-voltage (zero velocity) current range; the hysteresis stems from the already mentioned width-dependent fluxon potential and is characterized by a trapping current, i.e., the minimum current at which a fluxon still moves along the system, not being trapped by the potential and a depinning current, i.e., the maximum current at which the fluxon is pinned in the potential well. Furthermore, the step profile is not smooth but shows some fine structures due to the resonance of the traveling fluxon with wavelets radiated by the fluxon itself subject to a periodic width-dependent potential \cite{barbara}. As indicated by the premature step switching, this radiation prevents the fluxon from reaching relativistic speeds.

\subsection{Fluxon acceleration}
\vskip -10pt
\noindent Numerical simulations also showed that the fluxon (tangential) speed $\hat{u}\equiv d\hat{s_0}/d{\hat {t}} =(c/\lambda_J)\mathcal{Q}(\tau_0) d\tau_0/d{\hat{t}}$ is not constant. This is expected considering that a CAJTJ, by definition, does not have a constant width \cite{pagano,Benabdallah}; indeed, at variance with constant width junctions, the fluxon velocity cannot be determined by the balance between the driving force on the fluxon and the drag force due to the dissipative losses \cite{scott}. It was found that $\hat{u}$ is largest at the equatorial points, where the annulus is narrowest. More specifically, we found that, in absence of external magnetic field, $d\tau_0/d{\hat{t}}= \phi_{\hat{t}}(\tau_0,\hat{t}) /\phi_\tau(\tau_0,\hat{t}) \propto \mathcal{Q}^{-2}(\tau_0)$, i.e., $\hat{u}=\hat{u_0} \cosh\bar{\nu}/\mathcal{Q}(\tau_0)$, where $\hat{u_0}=\hat{u}(\tau_0=0)$ is a constant \cite{note}. It is interesting, at this point, to calculate the fluxon normal acceleration, which causes radiative losses \cite{scott}. The fluxon speed is tangential, ${\bf{\hat{u}}}=\hat{u}{\bf{\hat{T}}}$. Then the acceleration is ${\bf{\hat{a}}}= d{\bf{\hat{u}}}/d\hat{t}=(\lambda_J/c)\hat{u_0}^2\! \cosh^2\!\bar{\nu}\!\left[ \cosh\!\bar{\nu} \sin\!\tau_0 \left(\cos^2\!\tau_0-\sinh^2\!\bar{\nu} \right),\!-\sinh\!\bar{\nu} \cos\!\tau_0 \left(\sin^2\!\tau_0+\cosh^2\!\bar{\nu} \right) \right]\! /\mathcal{Q}^6(\tau_0)$. At last, the normal fluxon acceleration is $\hat{a}_\nu(\tau)\equiv {\bf{\hat{a}}} \cdot {\bf{\hat{N}}}= -(\lambda_J/c) \hat{u_0}^2 \cosh^2\bar{\nu} \sinh2\nu /2\mathcal{Q}^5(\tau_0)$. This acceleration, and so the radiation emitted by the fluxon, is largest where the curvature radius is smallest, i.e., at the ellipse's poles; in fact, at $\tau=\pm \pi/2$, we have $\mid a_\nu \mid=(\lambda_J/c) \hat{u_0}^2 \cosh^3\bar{\nu}/\sinh^4\bar{\nu}$. This indicates that very eccentric CAJTJs emit more radiation at the extremity as it occurs in linear \Jos transmission lines; in fact, for $\bar{\nu}<<1$ it is $\mid a_\nu \mid \approx (\lambda_J/c) \hat{u_0}^2 \bar{\nu}^{-4}$. The radiation frequency can be finely tuned by the bias current. Furthermore, the frequency range and the radiation power can be increased by inserting a larger number of fluxons in the annular junction. We note that no external field is needed for the operation of such an oscillator, at the variance with the flux-flow oscillator \cite{nagatsuma83}. The tangential fluxon acceleration is $\hat{a}_\tau(\tau)\equiv {\bf{\hat{a}}} \cdot {\bf{\hat{T}}}= (\lambda_J/c) \hat{u_0}^2 \cosh^2\bar{\nu} \sin2\tau/2\mathcal{Q}^5(\tau_0)$. 

\section{Conclusions}
\vskip -10pt
In this work we focused on the the static and dynamic properties of a not simply connected planar \Jos tunnel \jun shaped as an annulus delimited by two confocal ellipses, i.e., with a periodically varying width. We found that the nonlinear phenomenology of CAJTJs is richer than that of elliptic annular junction of constant width \cite{SUST15,JLTP16}, providing an elegant example of how the geometrical subtleties are of paramount importance in the physics of \Jos tunnel junctions. More specifically, it was found that the \Jos phase obeys to a modified and perturbed sine-Gordon equation that, although not integrable, admits (numerically computed) solitonic solutions. The key ingredient of this partial differential equation is an effective Josephson penetration length inversely proportional to the local \jun width. This spatial dependence, in turn, generates a periodic zero-mean potential that alternately attracts and repels the fluxons (or antifluxons). The fluxon tangential speed is not constant, even in the absence of an external magnetic field, and its mean value cannot be determined by the balance between the driving and drag forces. Indeed, a fluxon traveling around a long CAJTJ undergoes an inward acceleration much larger that in a uniform elliptic motion. This acceleration is associated with a radio-frequency power emission mainly concentrated at the ellipse equatorial points. 

\noindent More analytical considerations on Eq.(\ref{psge4}) as well as a thorough numerical investigation of the fluxon dynamics in CAJTJs will be the subject of a future search. A magnetic field applied in the junction plane gives rise to a tunable non-sinusoidal periodic potentials lacking spatial reflection symmetry and strongly dependent on the annulus eccentricity. The possibility to engineering magnetic double well potentials for the realization of robust Josephson vortex qubits also worth to be explored. 

\renewcommand{\theequation}{A-\arabic{equation}}
\setcounter{equation}{0}  
\setcounter{subsection}{0}  
\section*{Appendix A - The magnetic diffraction pattern of an \elli planar \Jos tunnel junction} 

Let us consider a planar \Jos tunnel \jun delimited by an ellipse of principal semi-axes $a=c\cosh\bar{\nu}<<\lambda_J$ and $b=2c\sinh\bar{\nu}$ in presence of a spatially homogeneous in-plane magnetic field applied at an arbitrary angle ${\bar{\theta}}$ with the $Y$-axis. The most general expression for the \Jos phase is (see Eq.(\ref{small})):

\begin{equation}
\phi(\nu,\tau,\bar{\theta})=h(\nu,\bar{\theta}) \sin[\xi(\nu,\bar{\theta})-\tau] + n\tau + \phi_0 
\label{small2}
\end{equation}

\noindent where $n$ is an integer number, called the winding number and  $\phi_0$ is an integration constant. The term $n\tau$ stems from the fact that the Josephson current density is an observable quantity and, according to Eq.(\ref{jj}), must be a single valued upon any closed path. For a simply connected junction any value of the winding number is, in principle, possible, but the state with $n=0$ is energetically preferred. However, for \ann junctions, $n$ corresponds to algebraic sum of \Jos vortices (or fluxons) trapped in the \jun due to flux quantization in one of the superconducting electrodes. Inserting Eq.(\ref{small2}) in Eq.(\ref{IJ}) and recalling that in elliptic coordinates the elementary surface can be written as $dS=\frac{1}{2}c^2 (\cosh\!2\nu + \cos \!2\tau)d\nu d\tau$, we get:
\vskip -10pt
\begin{equation}
I_J(h,\bar{\nu},\phi_0)\!=\!J_c c^2\!\int_0^{\bar{\nu}} \Im(\nu) d\nu,
\label{AZ}
\end{equation} 
\noindent where:
\vskip -15pt
\begin{equation}
\Im(\nu,\phi_0)=[\Im_{sc}(\nu) + \Im_{cc}(\nu)]\cos\phi_0 + [\Im_{cs}(\nu)+ \Im_{ss}(\nu)] \sin\phi_0
\label{AY}
\end{equation} 
\noindent and:
\vskip -15pt
$$\Im_{sc}(\nu)= \frac{\cosh\!2\nu}{2} \int_{-\pi}^\pi \sin[h \sin(\xi-\tau) + n\tau]\,d\tau =\frac{\cosh\!2\nu}{2} \int_{-\pi}^\pi \sin[-h \sin\tau + n(\tau-\xi)]\,d\tau ,$$
\vskip -15pt
$$\Im_{cc}(\nu)= \frac{1}{2}\int_{-\pi}^\pi \!\cos\!2\tau \sin[h \sin(\xi-\tau) + n\tau]\,d\tau = \frac{1}{2}\int_{-\pi}^\pi \!\cos\!2(\tau-\xi) \sin[-h \sin\tau + n(\tau-\xi)]\,d\tau,$$
\vskip -15pt
$$\Im_{cs}(\nu)= \frac{\cosh\!2\nu}{2} \int_{-\pi}^\pi \cos[h \sin(\xi-\tau) + n\tau]\, d\tau =\frac{\cosh\!2\nu}{2} \int_{-\pi}^\pi \cos[-h\sin\tau + n(\tau-\xi)]\, d\tau ,$$
\vskip -15pt
$$\Im_{ss}(\nu)= \frac{1}{2}\int_{-\pi}^\pi \!\cos\!2\tau \cos[h \sin(\xi-\tau) + n\tau]\, d\tau = \frac{1}{2}\int_{-\pi}^\pi \!\cos\!2(\tau-\xi) \cos[-h\sin\tau + n(\tau-\xi)]\, d\tau. $$
\noindent Being $h$ independent on $\tau$, we have:
\vskip -5pt
$$\Im_{sc}(\nu)=\frac{\cosh\!2\nu}{2} \left[\cos n\xi \int_{-\pi}^\pi \sin(-h \sin\tau + n\tau)\,d\tau - \sin n\xi \int_{-\pi}^\pi \cos(-h \sin\tau + n\tau)\,d\tau\right] =$$ $$=\frac{\cosh\!2\nu}{2} \left[ - \sin n\xi \int_{-\pi}^\pi \cos(-h \sin\tau + n\tau)\,d\tau\right]= -\pi \cosh\!2\nu \sin n\xi J_{n}(h);$$
\vskip -5pt
$$\Im_{cc}(\nu)\!=\! \frac{1}{2}\left[\cos n\xi \!\!\int_{-\pi}^\pi \!\!\!\!\!\!\cos\!2(\tau\!-\!\xi)\!\sin(\!-h \sin\tau \!+\! n\tau)\,d\tau \!-\! \sin n\xi \!\!\int_{-\pi}^\pi \!\!\!\!\!\!\cos\!2(\tau\!-\!\xi)\!\cos(h \sin\tau \!-\! n\tau)\, d\tau\right]\!=$$
$$= \pi \cos n\xi\, \sin2\xi \frac{J_{n-2}(h) - J_{n+2}(h)}{2} - \pi \sin n\xi\, \cos2\xi \frac{J_{n-2}(h) + J_{n+2}(h)}{2};$$
\vskip -5pt
$$\Im_{cs}(\nu)=\frac{\cosh\!2\nu}{2} \left[\cos n\xi \int_{-\pi}^\pi \cos(-h\sin\tau + n\tau)\, d\tau + \sin n\xi \int_{-\pi}^\pi \sin(-h\sin\tau + n\tau)\, d\tau\right]=$$
$$=\frac{\cosh\!2\nu}{2} \left[\cos n\xi \int_{-\pi}^\pi \cos(-h\sin\tau + n\tau)\, d\tau \right]=\pi \cosh\!2\nu \cos n\xi J_{n}(h);$$
\vskip -5pt
$$\Im_{ss}(\nu)\!=\!\frac{1}{2}\left[\cos n\xi \!\!\int_{-\pi}^\pi \!\!\!\!\!\!\cos\!2(\tau\!-\!\xi) \cos(h\sin\tau \!-\! n\tau)\, d\tau \!+\! \sin n\xi \!\!\int_{-\pi}^\pi \!\!\!\!\!\!\cos\!2(\tau\!-\!\xi) \sin(\!-\!h\sin\tau \!+\! n\tau)\, d\tau\right]\!=$$
$$=\pi \cos n\xi\, \cos2\xi \frac{J_{n-2}(h) + J_{n+2}(h)}{2} - \pi \sin n\xi\,\sin2\xi \frac{J_{n-2}(h) - J_{n+2}(h)}{2}.$$
\vskip -5pt
\noindent Therefore,
\vskip -15pt
$$\Im(\nu,\phi_0)=-\pi \cosh\!2\nu \cos\phi_0 \sin n\xi J_{n}(h)+$$
$$+ \pi \cos\phi_0 \cos n\xi\, \sin2\xi \frac{J_{n-2}(h) - J_{n+2}(h)}{2} - \pi \cos\phi_0 \sin n\xi\, \cos2\xi \frac{J_{n-2}(h) + J_{n+2}(h)}{2}+$$
$$ + \pi \cosh\!2\nu \sin\phi_0 \cos n\xi J_{n}(h)+$$
$$+\pi \sin\phi_0 \cos n\xi\, \cos2\xi \frac{J_{n-2}(h) + J_{n+2}(h)}{2} - \pi \sin\phi_0 \sin n\xi\,\sin2\xi \frac{J_{n-2}(h) - J_{n+2}(h)}{2}=$$
\begin{equation}
=\pi \sin(\phi_0-n\xi)\left[ \sin2\xi \frac{J_{n-2}(h) - J_{n+2}(h)}{2} + \cosh\!2\nu  J_{n}(h) + \cos2\xi \frac{J_{n-2}(h) + J_{n+2}(h)}{2}  \right],
\label{Inu}
\end{equation}
\vskip 5pt
\noindent with $\sin2\xi={\sin2\bar{\theta}\sinh2\nu}/(\cosh2\nu+\cos2\bar{\theta})$ and  $\cos2\xi=(1+\cos2\bar{\theta}\cosh2\nu)/(\cosh2\nu+\cos2\bar{\theta})$.
\vskip 5pt
\noindent $\Im(\nu)$ changes its sign for negative odd integers. For large field values one can use the asymptotic forms for the Bessel functions, $J_n(h)\approx \sqrt{2/\pi h} \cos (h-n\pi/2-\pi/4)$, observing that $J_{n-2}(h)$ and $J_{n+2}(h)$ oscillate in phase, while both are out of phase with respect to $J_{n}(h)$. The analytic primitive of $\Im(\nu)$ exists only for $n=0$, i.e., for $\Im_0(\nu)=\pi \sin\phi_0 \left[\cosh\!2\nu\, J_{0}(h) + \cos2\xi\, J_{2}(h) \right]$; in fact, after some lengthy algebraic manipulations, exploiting the identity $dh/d\nu =h\sinh2\nu/2q^2$, it is possible to verify that:
\vskip -10pt
\begin{equation}
\pi \sin\phi_0 \frac{d}{d\nu}\left[\sinh2\nu \frac{J_1(h)}{h} \right]=\Im_0(\nu).
\label{prim}
\end{equation}
\noindent Inserting Eq.(\ref{prim}) in Eq.(\ref{AZ}) we have (for $n=0$):
\vskip -10pt
\begin{equation}
I_J(h,\bar{\nu},\phi_0)\!=\pi \sin\phi_0 J_c c^2 \sinh2\bar{\nu}\, \frac{J_1(h)}{h}.
\label{AZ2}
\end{equation}


\begin{thebibliography}{99}
\bibitem{note1} Some useful properties of the hyperbolic functions are: $\sinh(\nu - \hat{\imath}\pi/2)= -\hat{\imath}\cosh\nu$, $\cos(\nu - \hat{\imath}\pi/2)= -\hat{\imath}\sinh\nu$, $\sinh (2\nu - \hat{\imath}\pi)= -\sinh\nu$, and $\cos(2\nu - \hat{\imath}\pi)= -\cosh\nu$.
\end{thebibliography}
\end{document}